\documentclass[
 reprint,
superscriptaddress,
 amsmath,amssymb,
 aps,
]{revtex4-1}
\usepackage{dcolumn}
\usepackage{bm}
\usepackage{tabularx}
\usepackage{hyperref}
\usepackage{cleveref}

\usepackage{amsmath,amssymb,amsfonts,bbm,graphicx}
\usepackage[latin1]{inputenc}
\usepackage[T1]{fontenc}
\usepackage{verbatim}
\usepackage{textcomp}
\usepackage{graphicx}
\usepackage{amsmath}

\usepackage{graphicx}
\usepackage{dcolumn}
\usepackage{bm}
\usepackage{xcolor}

\begin{document}

\preprint{APS/123-QED}

\title{Bi$_{2}$W$_{2}$O$_{9}$: a potentially antiferroelectric Aurivillius phase}

\author{H. Djani}
\affiliation{Centre de D\'eveloppement des Technologies Avanc\'ees, Cit\'e 20 ao\^ut 1956, Baba Hassen, Alger, Algeria}
\author{E. E. McCabe}
\affiliation{School of Physical Sciences, University of Kent, Canterbury, Kent, CT2 7NH, U.K.}
\author{W. Zhang}
\affiliation{Department of Chemistry, University of Houston, 112 Fleming Building, Houston, Texas 77204-5003, United States}
\author{P. S. Halasyamani}
\affiliation{Department of Chemistry, University of Houston, 112 Fleming Building, Houston, Texas 77204-5003, United States}
\author{A. Feteira}
\affiliation{Department of Engineering and Mathematics, Sheffield Hallam University, Sheffield, S1 1WB, U.K.}
\author{J. Bieder}
\affiliation{Theoretical Materials Physics, Q-MAT, CESAM,, Universit\'e de Li\`ege, All\'ee 6 ao\^ut, 17, B-4000, Sart Tilman, Belgium}
\author{E. Bousquet}
\affiliation{Theoretical Materials Physics, Q-MAT, CESAM,, Universit\'e de Li\`ege, All\'ee 6 ao\^ut, 17, B-4000, Sart Tilman, Belgium}
\author{Ph. Ghosez}
\affiliation{Theoretical Materials Physics, Q-MAT, CESAM,, Universit\'e de Li\`ege, All\'ee 6 ao\^ut, 17, B-4000, Sart Tilman, Belgium}

\date{\today}

\begin{abstract}
Ferroelectric tungsten-based Aurivillius oxides are naturally stable superlattice structures, in which A-site deficient perovskite blocks [W$_n$O$_{3n+1}$]$^{-2}$ ($n=1,2,3,...$) interleave with fluorite-like bismuth oxide layers [Bi$_2$O$_2$]$^{+2}$ along the $c$-axis. In the $n=2$ Bi$_2$W$_2$O$_9$ phase, an in-plane antipolar distortion  dominates but there has been controversy as to the ground state symmetry. Here we show, using a combination of first-principles density functional theory calculations and experiments, that the ground state is a non-polar phase of $Pnab$ symmetry. We explore the energetics of metastable phases and the potential for antiferroelectricity in this $n=2$ Aurivillius phase.

\end{abstract}

\maketitle

\maketitle
\section{Introduction}

Ferroelectric oxides with naturally layered perovskite-like structures are the subject of intensive research  owing to their technological importance and to the interplay between competing structural instabilities that can give rise to complex phase transition scenarii~\cite{benedek-djani}.
Aurivillius compounds constitute a family of such layered perovskite related-materials in which a fluorite-like [Bi$_2$O$_2$]$^{+2}$ layers alternate along the [001] stacking direction with perovskite [$A_{n-1}B_nO_{3n+1}$]$^{-2}$ blocks, with $n$ being the number of layers of $B$O$_6$ octahedra in the perovskite block. Related Families of layered perovskite-related materials include the Dion-Jacobson (DJ; of general formula $A'A_{n-1}B_{n}O_{3n+1}$) and Ruddlesden-Popper families (RP; of general formula $A_{n+1}B_{n}O_{3n+1}$).
The RP and DJ phases have gained significant interest of late due to the appearance of non-centrosymmetric polar structures driven by the coupling of non-polar structural distortions (including rotations of $B$O$_{6}$ octahedra).~\cite{benedek-prl, benedek-DJ} In contrast, the Aurivillius phases have been known for some time as "proper" ferroelectrics in which the primary order parameter describes in-plane polar displacements.~\cite{Hervoches_Lightfoot_1999, Bi3TiW2O9}.

The Aurivillius series Bi$_2$W$_n$O$_{3n+3}$ consists of structures composed of perovskite blocks of corner-linked WO$_6$ octahedra $n$ layers thick, with no $A$ cations in these perovskite blocks (i.e. layered analogues to WO$_3$). These perovskite blocks are separated by fluorite-like bismuth
oxide layers. Their ideal or aristotype structures are of tetragonal $I4/mmm$ symmetry but as for other Aurivillius phases, numerous structural distortions (including tilts of WO$_6$ octahedra and cation displacements) are possible.

\begin{figure}[t]
\centering\includegraphics[angle=0, scale=0.30]{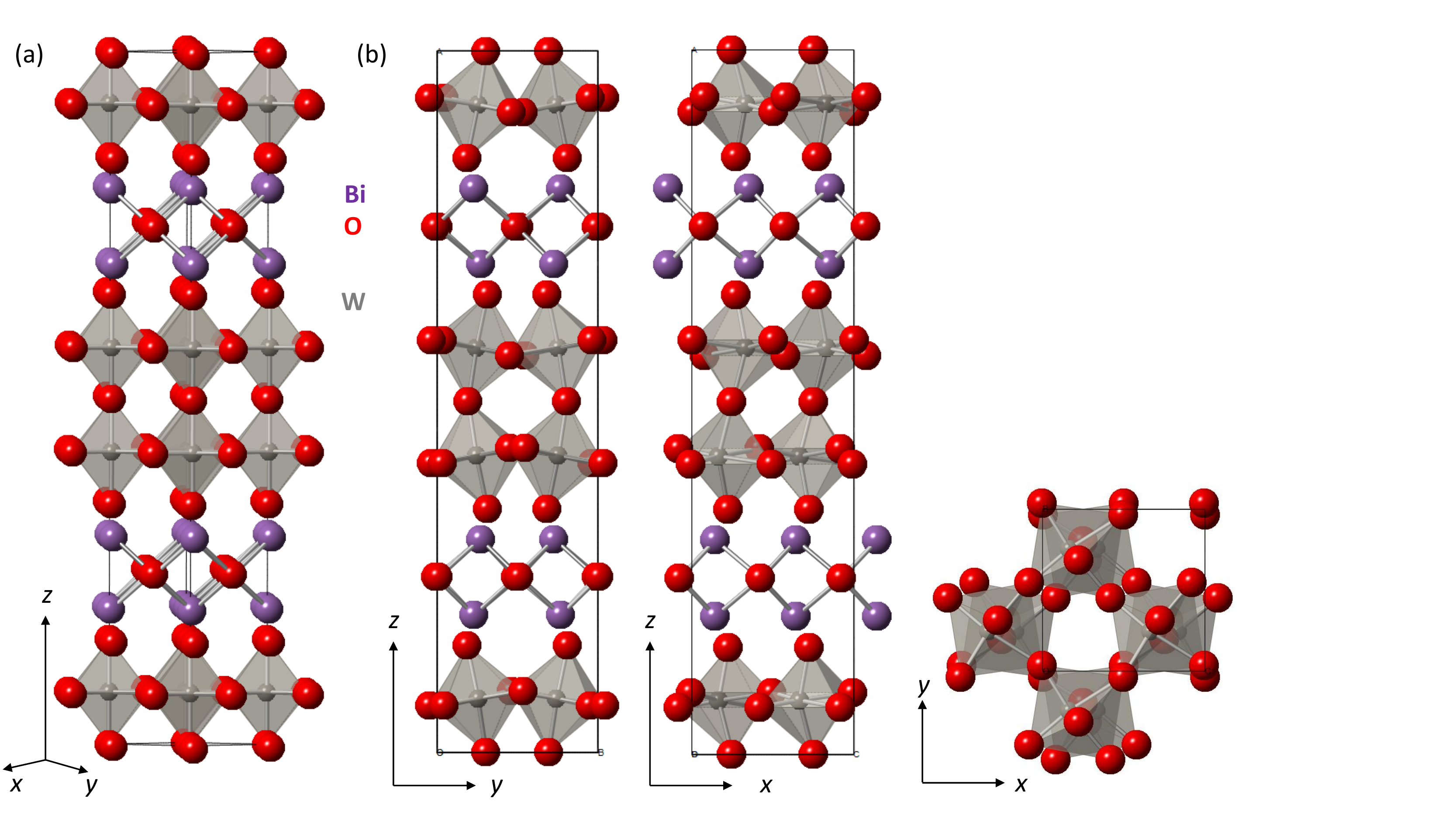}\\
\caption{The structure of Bi$_{2}$W$_{2}$O$_{9}$ in (a) the aristotype paraelectric  $I4/mmm$ phase and (b) the distorted  orthorhombic $Pnab$ ground state showing rotation of WO$_6$ octahedra about in-plane and out-of-plane axes and in-plane antipolar displacements of W$^{6+}$ ions. Bi, W and O ions are shown in purple, grey and red, respectively, and corner-linked WO$_6$ octahedra in grey.}
\label{fig1}
\end{figure}


The $n=1$ member of the series, Bi$_2$WO$_6$, the mineral known as russellite, is ferroelectric at room temperature, adopting a polar crystal structure of $P$2$_{1}ab$ symmetry with in-plane polar displacements and octahedral tilts around both in-plane and out-of-plane axes.~\cite{djani12, withers91, perez-mato04, perez-mato08, Okudera}

The $n=2$ member of the series, Bi$_2$W$_2$O$_9$, was first reported in 1938.~\cite{champarnaud} Its structure was investigated by Bando et \textit{al.} in 1979 using electron diffraction analysis and high resolution electron microscopy. It was described in terms of in-plane antipolar displacement of cations and octahedral tilts about both in-plane and out-of-plane axes. Although a polar space group ($Pna2_1$) was assigned, Bi$_2$W$_2$O$_9$ was described as potentially antiferroelectric, with in-plane cation displacements that are antiparallel from layer to layer along [001] direction~\cite{bando}. This model of $Pna2_1$ symmetry also implies an out-of-plane polar displacement along [001] which is rarely observed in Aurivillius phases but has been accepted as the ground state structure of Bi$_2$W$_2$O$_9$ on the basis of X-ray diffraction studies.~\cite{champarnaud} 


Antiferroelectrics (AFEs) form a class of functional materials that adopt a non-polar ground state but can undergo a phase transition in applied electric field to a polar ferroelectric state, provided that the non-polar and the polar phases are close enough in energy ~\cite{rabe13}. Such a field-induced phase transition gives rise to peculiar double hysteresis P versus E loops, which are appealing for data storage applications.  
Given the scarcity of antiferroelectric materials, the suggestion that Bi$_2$W$_2$O$_9$ might be antiferroelectric warrants further investigations.  

Here, we report first-principles density functional theory (DFT) calculations and symmetry analysis to explore theoretically the Born-Oppenheimer energy landscape of Bi$_2$W$_2$O$_9$ as well as neutron powder diffraction (NPD), second harmonic generation (SHG) and dielectric polarization measurements to support the assignment of the ground state. Our study shows that Bi$_2$W$_2$O$_9$ adopts a non-polar ground state of $Pnab$ symmetry (as illustrated in FIG.~\ref{fig1}) involving only antipolar displacements and in-plane and out-of-plane octahedral tilts.  DFT calculations reveal the presence of a metastable polar phase of $A2_1am$ symmetry (a common ground state of $n$=2 Aurivillius phases ~\cite{boullay12, perez-mato04}) only slightly higher in energy than the non-polar ground state. This metastable polar phase is consistent with  AFE behavior which unfortunately could not be accessed experimentally. During preparation of this manuscript we became aware of the single crystal X-ray diffraction study on Bi$_2$W$_2$O$_9$ which also proposes a ground state of $Pnab$ symmetry \cite{tian2018} fully consistent with our work.

\section{Methods}

\begin{figure*}[t]
\centering\includegraphics[angle=0, scale=0.50]{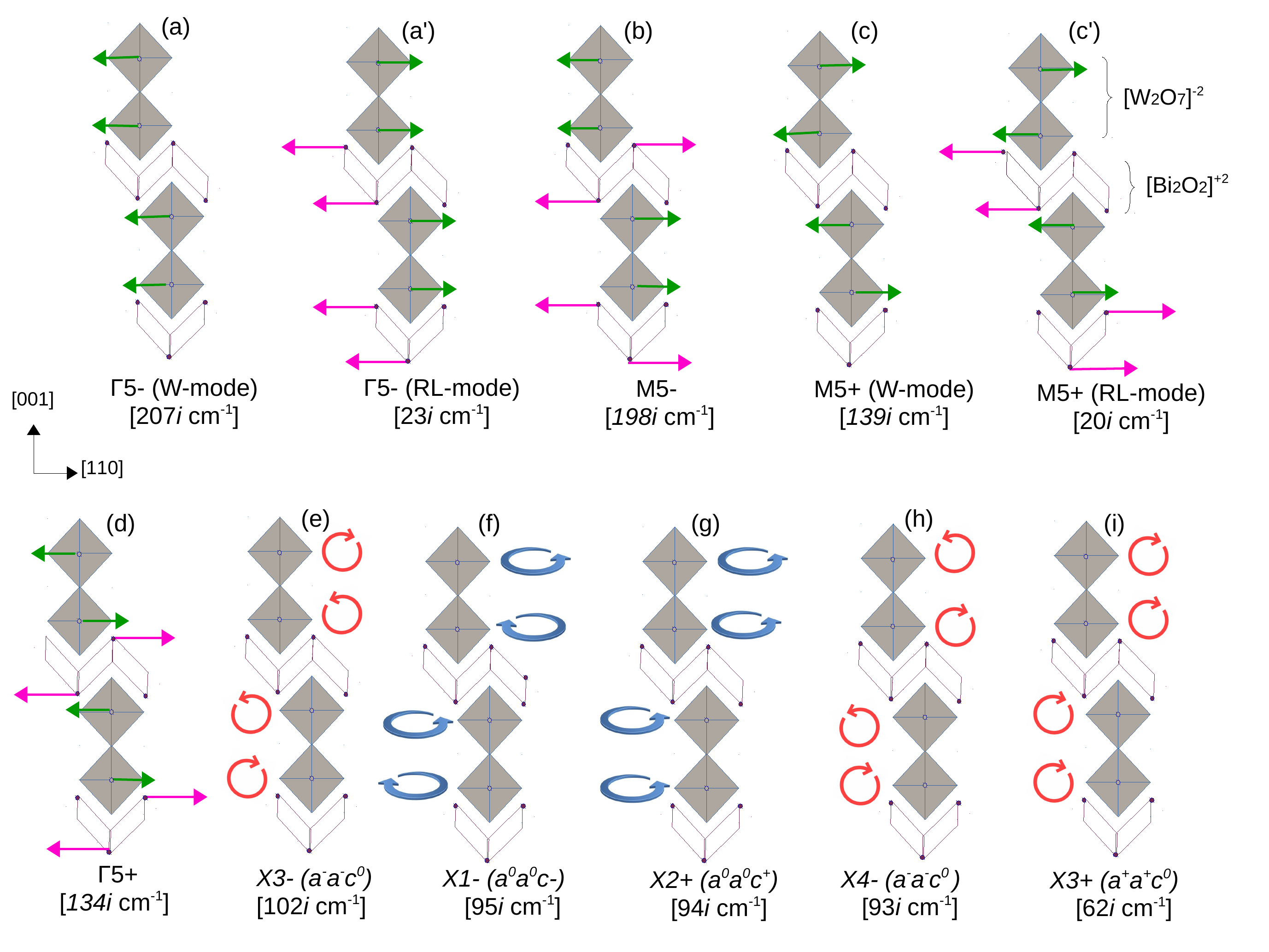}\\
\caption{Schematic illustration of atomic motions associated to the unstable phonon modes of the $I4/mmm$ phase of Bi$_2$W$_2$O$_9$ (at high-symmetry points of the Brillouin zone). Rigid-layer modes (RL modes) are related to a nearly rigid motion of the [Bi$_2$O$_2$]$^{+2}$ layer with respect to the perovskite block \cite{machado, djani12}. For modes involving polar and antipolar cationic displacements, oxygen atoms motions are omitted for clarity. The [110] direction is that in the tetragonal primitive cell and corresponds to the $a$-axis in the orthorhombic cell. 
\label{fig2}
}
\end{figure*}

\subsection{First-principles calculations}
Calculations were performed within DFT ~\cite{Hohenberg,KohnSham} using a plane waves method thanks to the ABINIT package ~\cite{abinit, abinit2, gonze09}. The exchange correlation energy functional was evaluated within  GGA PBEsol~\cite{Perdew2008} and  Bi (5d, 6s, 6p), W (5s, 5p, 5d, 6s) and O (2s, 2p) levels were treated as valence states in the norm-conserving pseudopotentials as delivered from the pseudodojo project \cite{dojo}. The wave functions were expanded up to a kinetic energy cutoff of 50 Hartrees. Integrals over the Brillouin zone were approximated by sums on a 6x6x1 Monkhorst-Pack mesh of special $k$-points. ~\cite{Monkhorst} The structural optimization was done using the Broyden-Fletcher-Goldfarb-Shanno minimization algorithm (BFGS) ~\cite{schlegel82}. We calculated the {\it ab initio} forces on the ions and relaxed the position of each atom until the absolute value of forces converged to less than 10$^{-5}$ Ha/Bohr. Phonons were calculated using Density Functional Perturbation Theory ~\cite{baroni01, gonze1997} and spontaneous polarization using the Berry phase formalism. ~\cite{resta94}

\subsection{Synthesis}
A polycrystalline sample of Bi$_{2}$W$_{2}$O$_{9}$ was prepared as a pale yellow powder by solid state reaction. Stoichiometric quantities of Bi$_{2}$O$_{3}$ (Alfa Aesar, 99.99$\%$ purity) and WO$_{3}$ (Sigma Aldrich, 99$\%$ purity) were ground together in an agate pestle and mortar and reacted in air at 750$^{\circ}$C for 12 hours and at 800$^{\circ}$C for 36 hours with intermittent grinding. 

\subsection{Characterisation}
Powder X-ray (XRPD) data were collected at room temperature using a PANalytical X'Pert3 powder diffractometer using Cu K$\alpha_{1}$  radiation, an X'Celerator detector and step size 0.04$^{\circ}$. Low temperature XRPD data were collected using an Oxford Cryosystems Phenix cryostat with the sample sprinkled onto a zero-background silicon wafer and $\sim$40 minute scans were collected on warming from 12 K to room temperature. Neutron powder diffraction (NPD) data were collected on the HRPD diffractometer (ISIS Neutron and Muon Source). A $\sim$8 g sample was loaded into a cylindrical vanadium can and data were collected at room temperature over 4 hours using both a 30 -- 130 ms window and a longer d-spacing 100 -- 200 ms window. Diffraction data were analysed using the Rietveld method ~\cite{Rietveld} using TopasAcademic software.~\cite{Coelho2003, Coelho2012} Combined X-ray and neutron (3 data banks) refinements were carried out, primarily using the 30 -- 130 ms NPD window. The background (shifted Chebyshev), zero point or sample height (DIFA/DIFC for neutron refinements), peak profiles, lattice parameters, atomic positions and isotropic thermal parameters were refined. Preferred orientation (using a March-Dollase function with a single preferred-orientation direction, consistent with a plate-like morphology for this top-loaded and pressed sample) was included to fit the XRPD data.~\cite{March, Dollase}. To check sample stoichiometry, the $Pnab$ model (see text) with a single global temperature factor was used for combined Rietveld refinement to refine fractional occupancies of Bi and O sites (while the W site occupancy was fixed at unity), this suggested occupancies close to unity for all sites (0.9886(7), 1.005(1), 0.997(3), 0.972(2), 0.978(2) and 0.987(2) for Bi, O(1), O(2), O(3), O(4) and O(5) sites, respectively), consistent with a composition close to stoichiometric. The web based ISODISTORT software ~\cite{isodistort} was used to explore possible structural distortions in terms of symmetry-adapted distortion modes. 

Bi$_{2}$W$_{2}$O$_{9}$ was tested for a second harmonic generation (SHG) signal using the experimental setup described in reference ~\cite{SHG}; a Bi$_{2}$W$_{2}$O$_{9}$ pellet was ground and sieved into distinct particle size ranges ($<$20, 20-45, 45-63, 63-75, 75-90, 90-125 $\mu$m). Relevant comparisons with known SHG materials were made by grinding and sieving crystalline KH$_{2}$PO$_{4}$ (KDP) into the same particle size ranges. SHG intensity was recorded for to different particle size ranges. No index matching fluid was used in any of the experiments.

Dielectric polarization measurements were carried out on sintered pellets of Bi$_{2}$W$_{2}$O$_{9}$: single-phase Bi$_{2}$W$_{2}$O$_{9}$ powders were pressed under an uniaxial applied load of 1 ton using a steel die with a diameter of 10 mm (Specac). These green bodies were subsequently fired in air at 860$^{\circ}$C for 2 hours to give dense ceramics suitable for high electric field measurements. Gold-sputtered electrodes were applied onto both faces of the fired ceramics and field-induced measurements were performed between -100$^{\circ}$C and 140$^{\circ}$C using an aixACCT system. Polarization (P) vs electric field (E) loops and leakage currents were recorded using a triangular signal at a frequency of 1 Hz. In addition, the P vs E response of ferroelectric Bi$_{2}$WO$_{6}$ ceramics was also measured for comparison.

\section{Results}
\subsection{First-principles calculations}
\label{DFT_results}


\begin{figure*}[t]
\centering\includegraphics[angle=0, scale=0.65]{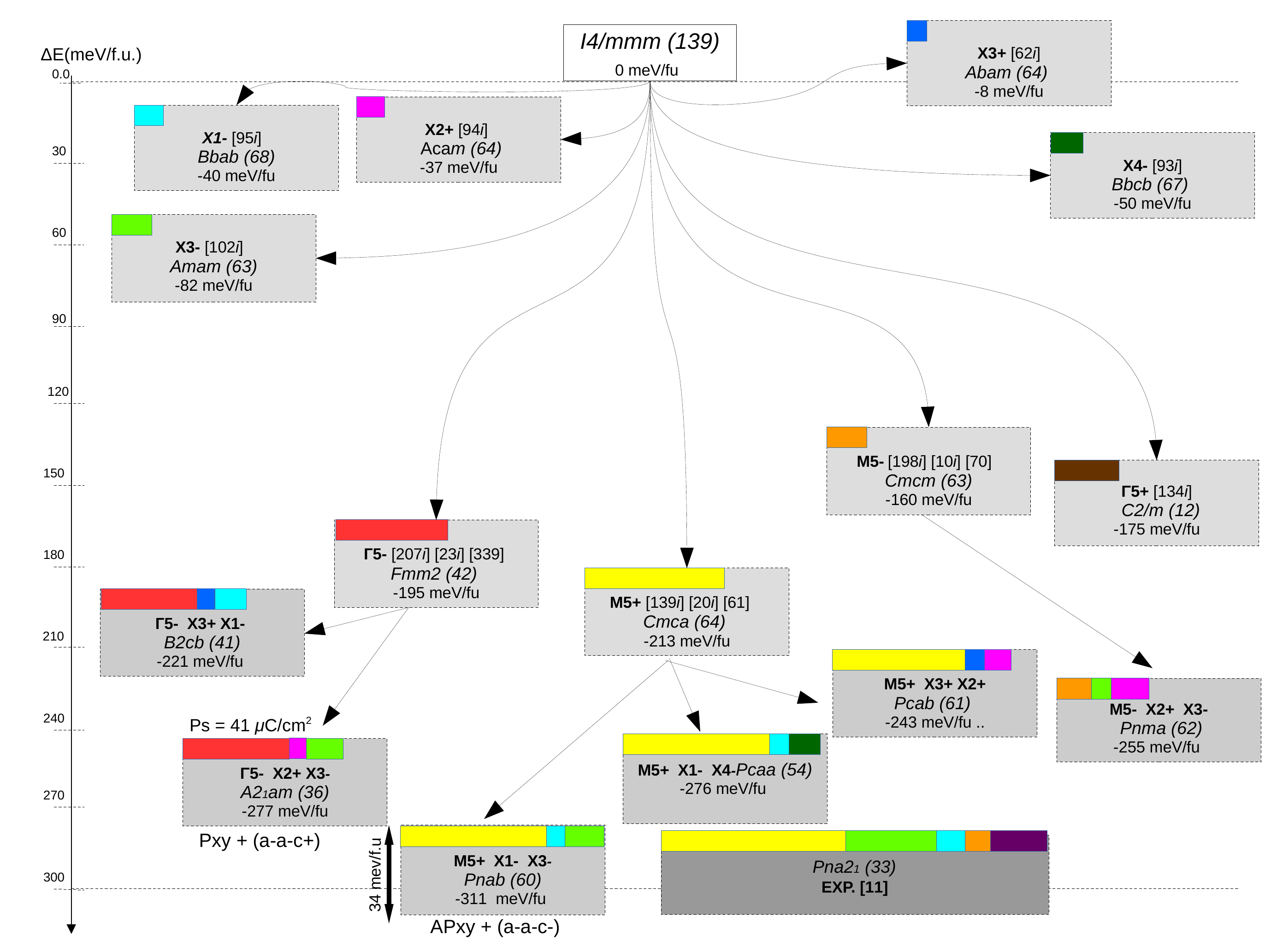}\\
\caption{Sketch of the most relevant metastable phases of Bi$_2$W$_2$O$_9$. Each phase is identified by the combination of modes giving rise to it, its symmetry and its energy ($\Delta$E in meV per formula unit) with respect to the $I4/mmm$ reference model.  Contributions of the modes of a given symmetry to the total atomic distortion of each phase with respect to the $I4/mmm$ reference model are identified through color segments, with lengths proportional to the projection ($A \alpha_i$, see main text) of these modes to the total distortion. A distinct color is affected to modes of distinct symmetry. When different modes of the same symmetry are contributing, only their total contribution is shown. The $Pna2_1$ phase from Ref. ~\cite{champarnaud}, although not a metastable phase in our computational framework, is included (darker grey box) for comparison.}
\label{fig3}
\end{figure*}

As mentioned in the introduction, the high-symmetry reference structure of Aurivillius compounds is  of $I4/mmm$ symmetry. To explore which combination of distortions can lower the energy and produce the ground state, phonon calculations are performed in the $I4/mmm$ paraelectric phase of  Bi$_{2}$W$_{2}$O$_{9}$. Numerous phonon  instabilities are identified at high-symmetry points, as illustrated in FIG.~\ref{fig2}. They include, on the one hand, in-plane atomic motions such as (a)-(a') $\Gamma_{5}^{-}$ polar motions (W- and Rigid-layer (RL)-modes), (b) M$_{5}^{-}$ inter-block antipolar motions, (c)-(c') M$_{5}^{+}$ intra-block antipolar motions (W- and RL-modes), and (d) $\Gamma_{5}^{+}$ intra-block antipolar motions  and, on the other hand, rotations of WO$_{6}$ octahedra about [001] and [110] axes, such as 
(e) X$_{3}^{-}$ rotation pattern $a_t^{-}a_t^{-}c^{0}$/$a_t^{-}a_t^{-}c^{0}$, 
(f) X$_{1}^{-}$ rotation pattern $a_t^{0}a_t^{0}c^{-}$/$a_t^{0}a_t^{0}c^{-}$, 
(g) X$_{2}^{+}$ rotation pattern   $a_t^{0}a_t^{0}c^{+}$/$a_t^{0}a_t^{0}c^{+}$,
(h) X$_{4}^{-}$ rotation pattern $a_t^{-}a_t^{-}c^{0}$/-($a_t^{-}a_t^{-}c^{0})$, 
(i) X$_{3}^{+}$  rotation pattern $a_t^{+}a_t^{+}c^{0}$/$a_t^{+}a_t^{+}c^{0}$ (with $t$ referring to the primitive tetragonal cell).


\begin{figure*}[t]
\centering\includegraphics[angle=0, scale=0.60 ]{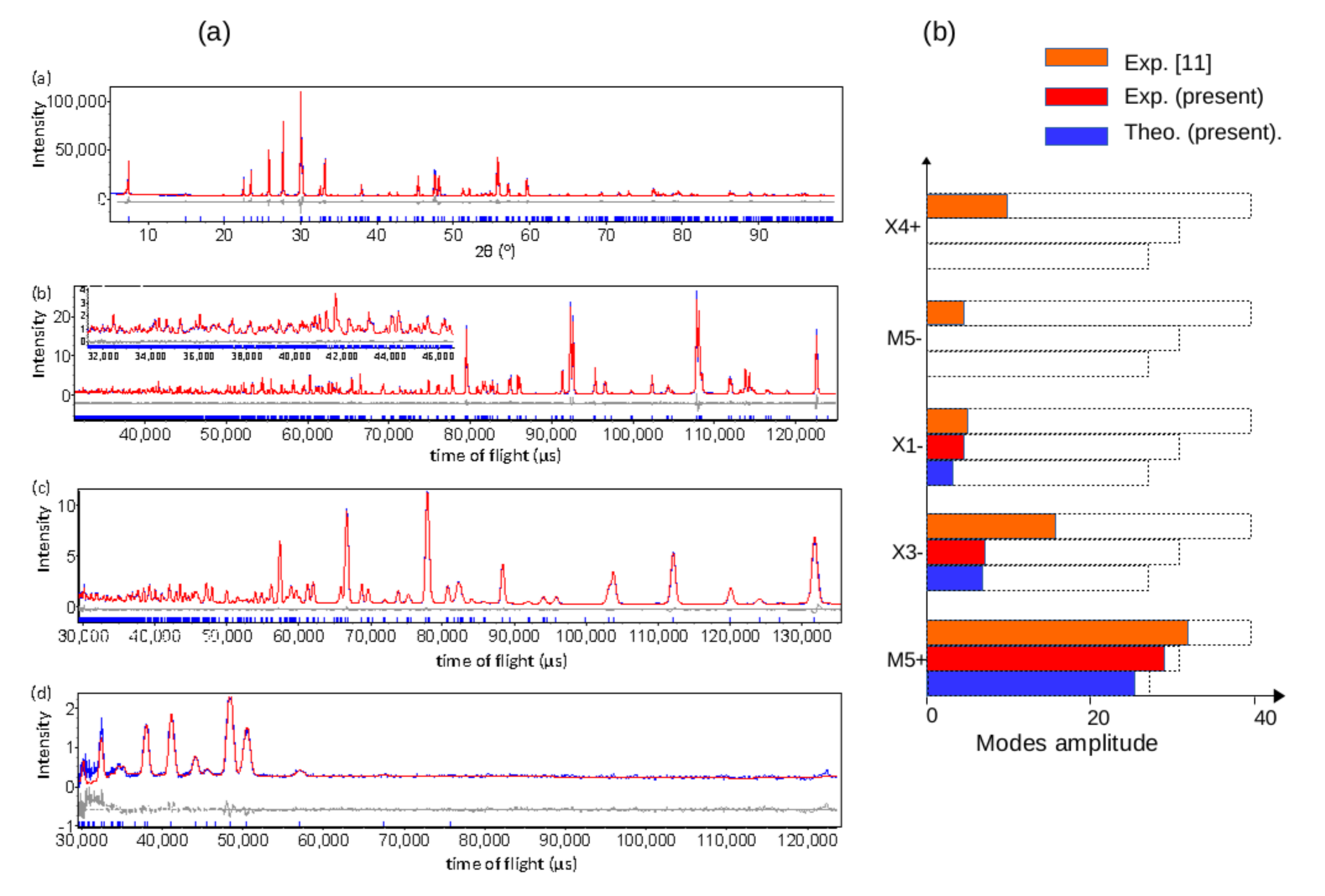}\\
\caption{\label{} (a) Rietveld refinement profiles using from combined refinement using (a) XRPD data, (b) backscattered 169$^{\circ}$ bank data ($\sim$0.6--2.6 \AA\ d-spacing range), including enlarged view of low d-spacing range, (c) 90$^{\circ}$ bank data ($\sim$0.8--3.8 \AA\ d-spacing range) and (d) 30$^{\circ}$ bank data ($\sim$2.3--9 \AA\ d-spacing range) collected for Bi$_{2}$W$_{2}$O$_{9}$ at room temperature using the $Pnab$ model. Observed and calculated (upper) and difference profiles are shown by blue, red and grey lines, respectively. (b) Decomposition of the full atomic distortion with respect to the $I4/mmm$ reference model for the $Pna2_1$ phase of Ref. ~\cite{champarnaud} (orange), the present $Pnab$ phase refined from our diffraction data (red) and the present $Pnab$ phase relaxed from first-principles calculations. The dashed boxes show the total distortion amplitudes ($A$) while the color segments show the respective contributions ($A \alpha_i$) of the modes of distinct symmetry.}
\label{fig4}
\end{figure*}


Condensation of these various instabilities (individually or together) into the reference $I4/mmm$ model, followed by full relaxation of atomic coordinates and cell parameters, allows one to identify a set of lower-energy (quantified by a negative $\Delta$E) metastable phases, as summarized in FIG.~\ref{fig3}.   

Considering first the metastable phases resulting from the condensation of individual unstable modes, the largest energy lowerings result from the M$_{5}^{+}$ mode consisting of antipolar motions within the perovskite block ($Cmca$ phase), followed by the polar $\Gamma_{5}^{-}$ displacement ($Fmm2$ phase). The antipolar modes M$_{5}^{-}$ ($Cmcm$ phase) (consisting of antipolar motion between perovskite blocks) and $\Gamma{5}^{+}$ ($C2/m$ phase) yield smaller but still sizeable energy lowerings. 
By contrast, distortions involving octahedral tilts produce significantly smaller energy lowerings with, in decreasing order, X$_{3}^{-}$ ($Amam$ phase, non standard setting of $Cmcm$), X$_{4}^{-}$ ($Bbcb$ phase, non standard setting of $Cmma$), X$_{1}^{-}$ ($Bbab$ phase, non standard setting of $Ccca$), X$_{2}^{+}$ ($Acam$ phase, non standard setting of $Cmca$) and X$_{3}^{+}$ ($Abam$ non standard setting of $Cmca$). This hierarchy corresponds to what is observed in WO$_3$, in which the polar and antipolar distortions lower the total energy much more than the octahedral tilts~\cite{hamdi}. This is in contrast with other Aurivillius phases like Bi$_2$WO$_6$, in which octahedral tilts and polar distortions give comparable energy gains (although the tilts are much less unstable than the polar mode). In SrBi$_2$Ta$_2$O$_9$, the strongest instability is the octahedral tilt X$_{3}^{-}$ mode which gives a similar energy lowering to the less unstable polar mode~\cite{PhysRevB.70.214111}.

Considering now the condensation of combinations of unstable modes, additional phases at lower energies can be identified. The most stable phases are those resulting from the condensation of  M$_{5}^{+}$ or $\Gamma_{5}^{-}$ displacements and $a^{-}a^{-}c^{0}$ (X$_{3}^{-}$ or X$_{4}^{-}$) tilts (i.e. models of $Pnab$, $Pcaa$ or $A$2$_{1}am$ symmetry). As noted by Zhang et al~\cite{La3Ni2X9}, the "sign" of rotation pattern around the in-plane axis from one perovskite block to the next  (i.e. $a_t^{-}a_t^{-}c^{0}$/+($a_t^{-}a_t^{-}c^{0}$ X$_{3}^{-})$ tilts, or  $a_t^{-}a_t^{-}c^{0}$/-($a_t^{-}a_t^{-}c^{0}$) X$_{4}^{-}$) tilts), changes the symmetry (and energy) of the resulting phase. 

\begin{figure*}[t] 
\includegraphics[width=0.95\linewidth]{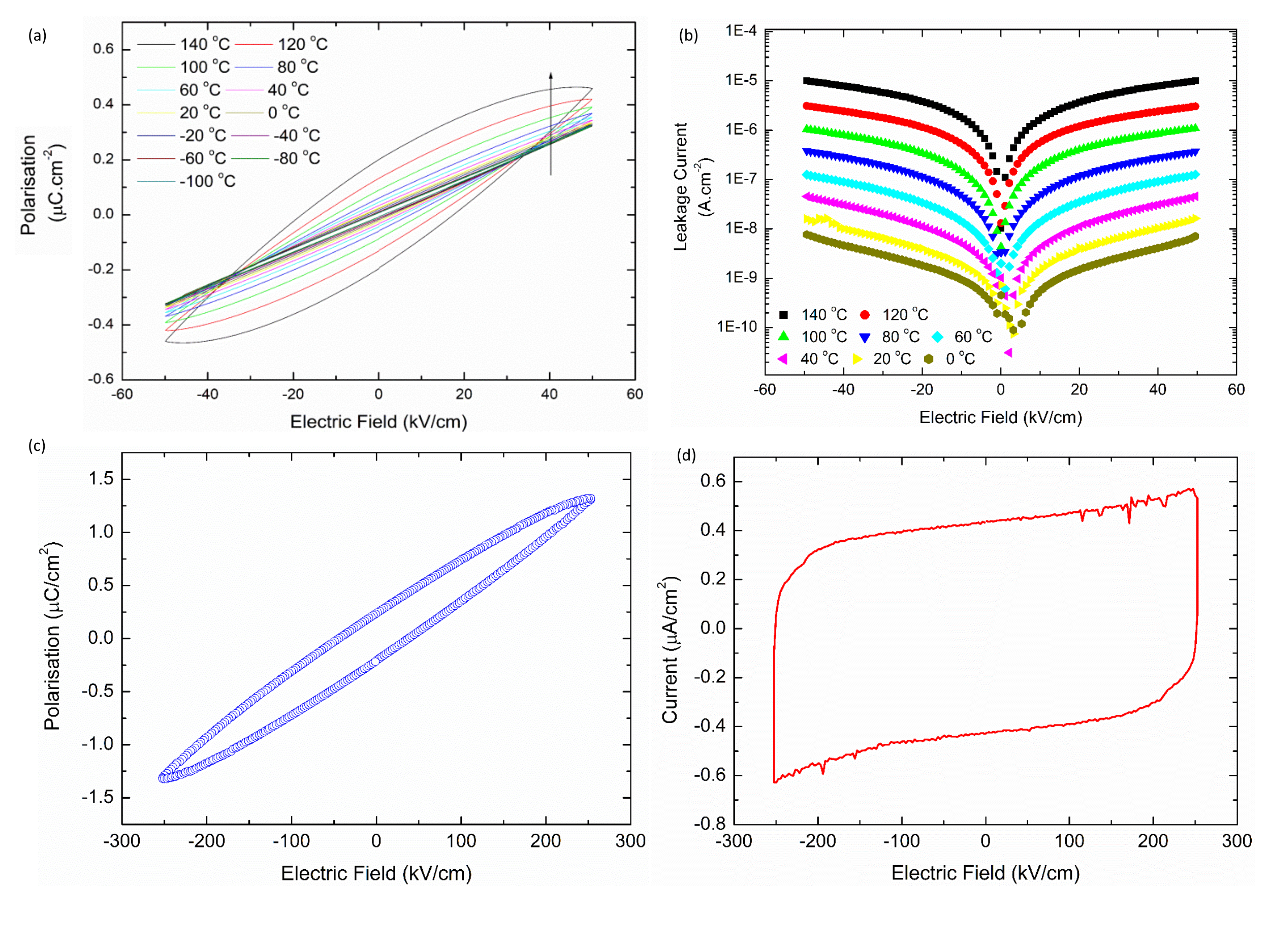}
\caption{[color online]  Dielectric measurements on sintered pellets of Bi$_{2}$W$_{2}$O$_{9}$, (a) showing polarisation and (b) showing leakage current with applied field at various temperatures; (c) and (d) show polarisation and current density for fields up to $\pm$250 kV cm$^{-1}$ at -40$^{\circ}$C.}
\label{fig5}
\end{figure*}

The combination of one of the in-plane polar/antipolar displacements ($\Gamma_{5}^{-}$/M$_{5}^{+}$) with one of the in-plane axis tilts (X$_{3}^{-}$, X$_{4}^{-}$ and X$_{3}^{+}$)  allows systematically the emergence of a second out-of-plane axis octahedral tilt (X$_{1}^{-}$ or X$_{2}^{+}$ tilts) by trilinear coupling. This is in contrast to Bi$_2$WO$_6$, for which in-plane and out-of-plane axis octahedral tilts (X$_{3}^{+}$ and  X$_{2}^{+}$ modes) combine with polar $\Gamma_{5}^{-}$ mode in the $P2_1ab$ ground state without the action of trilinear coupling~\cite{djani12}.


In order to characterize the atomic distortion $\Delta$ of each metastable phase with respect to the $I4/mmm$ tetragonal reference structure, we can express $\Delta$ in the  basis of atomic displacements formed by the phonon eigendisplacement vectors $\eta_i$ of the $I4/mmm$ phase (such that $<\eta_i| M | \eta_j> = \delta_{ij}$), following the scheme explained in Ref.~\cite{djani12}: $\Delta = A \sum_i \alpha_i \eta_i$, where $A$ is the total distortion amplitude and $\alpha_i$ are the relative mode contributions such that $\sum_i \alpha_i^2=1$. The contributions of distinct phonon modes $i$ to the distortion $\Delta$ of a given phase correspond therefore to the amplitudes $A \alpha_i$, illustrated in distinct colors in FIG.~\ref{fig3}.  

The condensation of an unstable mode of a given symmetry automatically allows the appearance of other unstable and stable modes of the same symmetry. In particular, the condensation of  $\Gamma_{5}^{-}$-W or M$_{5}^{+}$-W modes allows the related unstable RL-modes, as well as stable harder modes that have significant contributions to the total distortion. When different modes of the same symmetry appear together, only their global contribution is shown for simplicity in FIG.~\ref{fig3} \footnote{When distinct modes $j$ of a given symmetry contribute to the same distortion $\Delta$, their  total contribution corresponds to $A \sqrt{(\sum_j \alpha_j^2)}$} while detailed individual mode-by-mode contributions are reported in SI (Table SI11.1).

We have seen that the largest lowerings of energy are produced by the $M_5^+$ and $\Gamma_5^-$ distortions, the $Cmca$ phase being a little lower in energy than the $Fmm2$ phase. The additional condensation of $X_3^-$ in both these phases then brings the system to the $Pnab$ and $A2_1am$ phases with the appearance of a very similar third mode in each case ($X_1^-$ or $X_2^+$). As shown in FIG.~\ref{fig3}, the distortion amplitudes in the $Pnab$ and $A2_1am$ phases remain very similar to those in the phases with single distortions and the total energy lowerings are roughly comparable to the sum of the energy lowerings produced by the individual distortions. This highlights that all these distortions are only weakly coupled (i.e. they do not strongly compete nor cooperate). So, the $X_3^-$ distortion plays a very important role in further decreasing the energies of the $Fmm2$ and $Cmca$ phases but does not revert their relative stability.

From this, we identify the $Pnab$ phase, combining in-plane antipolar $M_5^+$ cation displacements, $X_1^-$ out-of-plane axis octahedral rotation  and $X_3^-$ in-plane axis octahedral tilt, as the ground state structure of Bi$_2$W$_2$O$_9$. Allowing additional condensation of the $\Gamma_{3}^{-}$ out-of-plane polar displacements does not provide any extra energy gain. It is noticeable that the $Pna2_{1}$ model reported in the literature ~\cite{bando,champarnaud,Maczka} has similar contributions from  M$_{5}^{+}$, X$_{3}^{-}$ and X$_{1}^{-}$ distortions, but involves additional X$_{4}^{+}$ and M$_{5}^{-}$ modes (see FIG.~\ref{fig4}(b)). These modes break the inversion symmetry, allowing the appearance of out-of-plane polar displacement $\Gamma_{3}^{-}$ thanks to trilinear couplings :  M$_{5}^{+}$ $\oplus$ M$_{5}^{-}$ $\oplus$ $\Gamma_{3}^{-}$ and X$_{3}^{-}$ $\oplus$ X$_{4}^{+}$ $\oplus$ $\Gamma_{3}^{-}$. A full structural relaxation of this literature model only preserves $M_5^+$, $X_1^-$ and $X_3^-$ distortions and suppresses  X$_{4}^{+}$ and M$_{5}^{-}$, consistent with the $Pnab$ ground state. Further phonon calculations in $Pnab$ show no remaining instability, ruling out the possibility to go from $Pnab$ to $Pna2_1$ by adding either X$_{4}^{+}$, M$_{5}^{-}$ or $\Gamma_{3}^{-}$.

Bi$_{2}$W$_{2}$O$_{9}$ might have been expected to adopt a polar $A$2$_{1}am$ ground state, in line with other stoichiometric $n$ = 2 Aurivillius phases. In Bi$_{2}$W$_{2}$O$_{9}$, such a polar phase is metastable and significantly stabilized with respect to the reference $I4/mmm$ phase but it appears 34 meV/f.u higher in energy than the $Pnab$ ground state. From Berry phase calculations, its spontaneous polarization is estimated to be 41 $\mu$C cm$^{-2}$ along the $a$-axis of the orthorhombic cell. Although Bi$_{2}$W$_{2}$O$_{9}$ is not ferroelectric, following Rabe \cite{AFE_Rabe},  the proximity in energy of this $A$2$_{1}am$ polar phase with the non-polar $Pnab$ ground state makes it a potential antiferroelectric material. Indeed, by applying an electric field, it might be possible to stabilize the $A$2$_{1}am$ phase against the $Pnab$ one and open a double hysteresis loop typical of an antiferroelectric. The field ${\cal E}_c$ required to make the $A$2$_{1}am$ phase thermodynamically more stable than the $Pnab$ phase  can be estimated by ${\cal E}_c$= $\Delta E$ / $\Omega_0 P_s$ where $\Delta E$ is the energy difference between the two phases (34 meV/f.u.), $P_s$ is the spontaneous polarization of the polar phase (41 $\mu$C cm$^{-2}$), and $\Omega_0$ is its unit-cell volume (690 \AA ), yielding the relatively modest value ${\cal E}_c = 192$ kV/cm.

\subsection{Neutron powder diffraction}
The main peaks in both XRPD and NPD data are consistent with a primitive, orthorhombic structure with unit cell $a \sim b \sim a_{t}\sqrt2 \sim 5.4$ \AA\  and $c \sim 23.7$ \AA\ ($t$ refers to a hypothetical tetragonal aristotype structure of $I4/mmm$ symmetry). The data could be fitted well with the reported non-centrosymmetric structure of $Pna$2$_{1}$ symmetry ~\cite{bando,Maczka,champarnaud} (see SI), although the polar displacement along the $c$ axis is very small and other possible structures as described in Section \ref{DFT_results}) were also considered (see SI).


The ISODISTORT software~\cite{isodistort} was used to consider possible distortions and the "mode inclusion method"~\cite{Tuxworth,La2O2Fe2OSe2} was employed, using the highest resolution backscattered (169$^{\circ}$) NPD data, to determine which distortions gave the greatest improvement in fit. This analysis indicated that in-plane antipolar displacements (M$_{5}^{+}$) and in-plane axis octahedral tilts $a^{-}a^{-}c^{0}$ (X$_{3}^{-}$) gave significant improvements in fit, together lowering the symmetry to $Pnab$. (As discussed above, these two distortions couple and also allow the out-of-plane axis octahedral $a^{0}a^{0}c^{-}$ rotation (X$_{1}^{-}$)). Allowing further distortions, such as the out-of-plane polar displacement ($\Gamma_{3}^{-}$) to give the reported $Pna$2$_{1}$ model did not give a significant improvement in fit. Other similar models, including the $Pcaa$ model described above, as well as models of $Pcab$, $Pnaa$ and $P$2$_{1}ab$ symmetry were also considered and could be discounted (see further discussion in SI).

A careful check was made for visible improvements in fit on moving from the non-polar $Pnab$ model to either polar $Pca$2$_{1}$ or $Pna$2$_{1}$ models and although additional reflections are allowed for both polar models, no intensity was observed in these reflection positions (see SI). Fitting statistics suggest a slight improvement in fit for the polar $Pna$2$_{1}$  model compared with the fit for the $Pnab$ model (R$_{wp}$s of  4.09$\%$ (134 parameters) and 4.13$\%$ (109) parameters for $Pna$2$_{1}$ and $Pnab$ models, respectively). Hamilton $t$ tests~\cite{Hamilton} were carried out using the fitting statistics for the highest resolution backscattered (169$^{\circ}$) NPD data and suggested that the improvement in fit, given the increased number of parameters, is significant at the 25$\%$ significance level (an $R$ factor ratio of 1.05 was obtained for comparison with the polar $Pna$2$_{1}$ model with the $Pnab$ model, compared with a calculated value of 1.01). However, this weighted residual method is known to be less reliable for differentiating between centrosymmetric and related non-centrosymmetric models for powder diffraction data~\cite{Whitaker} and, in the absence of any visible improvement in fit (or conclusive physical property measurements consistent with a polar phase, see below), the centrosymmetric $Pnab$ model is the most appropriate to describe the average crystal structure of Bi$_{2}$W$_{2}$O$_{9}$. Final refinement details and selected bond lengths are given in Tables \ref{Pbcn_params} and \ref{bond_lengths} and refinement profiles are shown in FIG.~\ref{fig4}(a). The structure is equivalent to what illustrated in FIG.~\ref{fig1}

\begin{table}[ht]
\caption{Details from Rietveld refinement using room temperature XRPD and NPD data for Bi$_{2}$W$_{2}$O$_{9}$ using $Pnab$ model with $a$ = 5.43349(7) \AA, $b$ = 5.41326(7) \AA, $c$ = 23.6902(3) \AA; $R_{wp}$ = 3.41$\%$, $R_{p}$ = 6.93$\%$, $\chi^{2}$ = 5.37 (109 parameters). }
\centering
\begin{tabular} {c c c c c c}
\hline\hline
Atom & site  & $x$ & $y$ & $z$ & $U_{iso}$\time100(\AA$^{2}$) \\
\hline\hline
Bi   & 8$d$ & 0.85422(9)  & 0.7298(1) & 0.69600(2) & 0.36(1) \\
W    & 8$d$ & 0.6657(1)   & 0.7449(2) & 0.07670(3) & 0.10(1) \\
O(1) & 8$d$ & -0.0752(1)  & 0.5099(1) & 0.25004(4) & 0.32(1) \\
O(2) & 4$c$ & 0.25        & 0.3055(2) & 0          & 0.48(2) \\
O(3) & 8$d$ & 0.0808(1)   & 0.4534(1) & 0.56642(3) & 0.64(2) \\
O(4) & 8$d$ & 0.5030(1)   & 0.0362(1) & 0.58779(3) & 0.53(1) \\
O(5) & 8$d$ & 0.7243(1)   & 0.8131(1) & 0.15224(4) & 0.56(2) \\
\hline
\label{Pbcn_params}
\end{tabular}
\end{table}

\begin{table}[ht]
\caption{Selected bond lengths from Rietveld refinement using room temperature XRPD and NPD data for Bi$_{2}$W$_{2}$O$_{9}$ using $Pnab$ model. Bond valence sum calculations suggest observed valences of 2.9 and 6.2 for Bi and W sites, respectively. }
\centering
\begin{tabular} {c c | c c}
\hline\hline
Bond & Length (\AA)  & Bond & Length (\AA) \\
\hline\hline
Bi---O(1) & 2.1816(9) & W---O(2) & 1.8936(7) \\
Bi---O(1) & 2.2506(9) & W---O(3) & 1.769(1) \\
Bi---O(1) & 2.3131(9) & W---O(3) & 2.150(1) \\
Bi---O(1) & 2.4974(9) & W---O(4) & 1.796(1) \\
Bi---O(5) & 2.5192(9) & W---O(4) & 2.141(1) \\
Bi---O(5) & 2.5244(8) & W---O(5) & 1.855(1) \\
\hline
\label{bond_lengths}
\end{tabular}
\end{table}

These diffraction data can only reveal the average, long-range crystal structure adopted by Bi$_{2}$W$_{2}$O$_{9}$; we note that bond valence sum calculations~\cite{BVS1,BVS2} give observed valences close to those expected for Bi$^{3+}$ sites, but that W$^{6+}$ sites are slightly overbonded (6.2--6.3) and we cannot rule out the possibility of short-range distortions (either polar or non-polar) that might relieve this overbonding at a more local level.

\subsection{SHG and Dielectric polarization measurements}
Second harmonic generation (SHG) measurements using 1064 nm radiation gave a low (0.2 times that of KDP) phase-matcheable signal, suggesting a non-centrosymmetric component to the sample (see SI). However, the threshold for laser damage of Bi$_{2}$W$_{2}$O$_{9}$ is low (the sample was visibly damaged by the laser with the off-white sample developing dark brown spots as shown in SI). Such laser damage has been shown to generate white light, including a green component, which could account for the low signal detected in these SHG measurements.~\cite{optical-breakdown} The results of these SHG measurements are therefore inconclusive as the SHG signal could arise from non-centrosymmetric decomposition products such as Bi$_{2}$WO$_{6}$~\cite{Bi2WO6} or simply from optical breakdown of the material.~\cite{optical-breakdown}

Dielectric polarization measurements were also carried out on sintered pellets of Bi$_{2}$W$_{2}$O$_{9}$. Care was taken not to introduce ferroelectric impurity phases during the sintering process, as noted in earlier studies.~\cite{Feteira} Polarisation vs electric field measurements carried out at a range of temperatures did not show any saturation in polarisation (and no peaks in current density were observed) (FIG.~\ref{fig5}) and indicate that Bi$_{2}$W$_{2}$O$_{9}$ becomes increasingly leaky on warming, in contrast to what was observed in ferroelectric Bi$_{2}$WO$_{6}$ (see SI). Applying very large fields across the pellet at -40$^{\circ}$C did not reveal any field-dependent behavior for electric fields up to $\pm$250 kV cm$^{-1}$. These property measurements give no evidence for ferroelectricity in Bi$_{2}$W$_{2}$O$_{9}$ and are consistent with the assignment of the non-polar, centrosymmetric $Pnab$ symmetry to Bi$_{2}$W$_{2}$O$_{9}$ at room temperature suggested by first principles calculations and NPD analysis as discussed above.

\section{Discussion}

The diffraction data are consistent with the $Pnab$ symmetry predicted from first-principles calculations and, as illustrated in FIG.~\ref{fig4}(b), refinements at room temperature  also yield an atomic structure and amplitudes of distortions in good quantitative agreement with the computations. Moreover, the variable temperature XRPD data show a smooth decrease in unit cell volume on cooling (see SI) with no evidence for discontinuities in lattice parameters that might indicate a low temperature phase transition to another phase (although additional low temperature NPD data would be necessary to confirm this), supporting the prediction that this $Pnab$ is the ground state. Our structure model is also consistent with that described by Bando {\it et al.} in term of antipolar displacements of W$^{6+}$ ions towards edges of the octahedra and a combination of tilts  $a^{-}a^{-}c^{-}$ ~\cite{bando} but suggests that the assignment of the polar $Pna2_1$ space group~\cite{champarnaud} is not correct. We find no evidence (from theory or experiment) for polarization along [001] and this is consistent with recent single crystal X-ray diffraction experiments \cite{tian2018}.


It is worth noting that the polar $A$2$_{1}am$ phase ($a^{-}a^{-}c^{+}$ tilts, with in-plane polar displacements) stated as the ground state of several other $n$ = 2 Aurivillius phases (e.g. Bi$_{2}A$Nb$_{2}$O$_{9}$ $A$ = Sr, Ca, Ba, Pb~\cite{Bi2ANb2O9}) has not been observed for Bi$_{2}$W$_{2}$O$_{9}$, but our calculations reveal that it is nevertheless a metastable phase with an energy only slightly above that of the non-polar $Pnab$ ground state. 

Bando {\it et al.} first suggested the possibility of antiferroelectricity in Bi$_{2}$W$_{2}$O$_{9}$. ~\cite{bando} Antiferroelectricity is not a property intrinsic to a specific structure but relates to the experimental ability to switch from a non-polar (antipolar) ground state to a polar phase under an applied electric field, opening up a double-hysteresis P versus E loop \cite{lines}. This was recently reinvestigated by Rabe~\cite{AFE_Rabe} who proposed practical conditions for a material to realize such a requirement: antiferroelectrics should show a ground state arising from a non-polar distortion of a high-symmetry reference phase and at the same time possess an alternative polar phase, appearing as a distortion of the same reference phase and sufficiently close in energy that an electric field can induce a first-order transition from the non-polar ground state to that polar phase. Guennou and Toledano~\cite{AFE_transitions} then built on this to clarify the symmetry requirements for antiferroelectricity~\cite{AFE_transitions}.  

The $Pnab$ and $A2_1am$ phases of Bi$_{2}$W$_{2}$O$_{9}$ satisfy the symmetry requirements of Guennou and Toledano and from the calculations, appear sufficiently close in energy for a predicted electric field of $\approx 200$ kV/cm to stabilize the polar phase over the ground state. Unfortunately, experimental measurements in fields up to 250 kV/cm were not able to realise this transition. This may result from materials science issues (eg. microstructure considerations).  It may also reflect the fact that the theoretical estimation of the switching field is too small or that the energy barrier between the two phases is large. This aspect warrants further investigation, in particular, strain engineering in thin films  might tune the energy landscape and decrease the energy difference between the two phases, eventually inverting the relative stability of the two phases.

Bi$_{2}$W$_{2}$O$_{9}$ shows many similarities with WO$_3$, which was also recently proposed to be potentially antiferroelectric \cite{hamdi2016} and it is worth noting that the energy landscape of both compounds share very similar features: the strongest instability is polar and produces, together with the antipolar instability, a more substantial gain in energy than octahedral rotations. However, the combination of antipolar motions and octahedral rotations yield a slightly larger gain in energy than polar motions, yielding a non polar ground state. The ground state of WO$_{3}$ involves a combination of antipolar distortions and octahedral rotations ($P2_1/c$ symmetry model allowing $a^{-}a^{-}c^{-}$ rotations and antipolar displacement~\cite{hamdi2016}) that is rather analogous to the pattern of displacements of the WO$_3$ bi-layer in the $Pnab$ phase of Bi$_{2}$W$_{2}$O$_{9}$. 
However, we note that: 1) the antipolar displacement in WO$_3$ is different from those of Bi$_{2}$W$_{2}$O$_{9}$; ie. the antipolar motion of W atoms in  WO$_3$ is between [001] columns (inside one column the displacement is polar see SI Fig. SI11.2) while  the antipolar motion of W atoms in Bi$_{2}$W$_{2}$O$_{9}$  is between [110]$_t$ layers (inside one layer the displacement is polar); 2) in bulk WO$_3$ combinations of antipolar distortions with either $a^{-}a^{-}c^{-}$ or $a^{-}a^{-}c^{+}$ give models of very similar energies in contrast to Bi$_{2}$W$_{2}$O$_{9}$ that give models of very distinctive energies (see $Pnma$ and $Pcca$ or $Pnab$ in FIG.~\ref{fig3}.

Aurivillius phases are analogous in term of symmetry to Ruddlesden-Popper compounds which are of potential interest as hybrid-improper ferroelectrics. It is interesting to compare Bi$_{2}$W$_{2}$O$_{9}$ with recently characterised Sr$_{3}$Zr$_{2}$O$_{7}$ and Sr$_{3}$Sn$_{2}$O$_{7}$.~\cite{Sr3Zr2O7,Sr3Sn2O7} 
Both have polar $A$2$_{1}am$ ($a^{-}a^{-}c^{+}$) and non-polar $Pnab$ ($a^{-}a^{-}c^{-}$) states very close in energy but in these $n$=2 Ruddlesden-Popper phases, it is the polar $A$2$_{1}am$ phase that is the ground state. 
Both phases undergo first-order phase transitions to $Pnab$ phases on warming. However, these Sr-containing Ruddlesden-Popper phases differ from Bi$_{2}$W$_{2}$O$_{9}$ in that the energy gains from octahedral tilts (e.g. X$_{1}^{-}$, X$_{3}^{-}$, X$_{2}^{+}$) are much greater than those from polar or antipolar displacements ($\Gamma{5}^{-}$ or M${5}^{+}$), whilst for Bi$_{2}$W$_{2}$O$_{9}$, these latter distortions are at least twice as favourable as octahedral tilts. This illustrates that the appearance of the polarization in the $A$2$_{1}am$ is more "proper" in Bi$_{2}$W$_{2}$O$_{9}$ and "hybrid-improper" in the Ruddlesden-Popper phases \cite{benedek-djani}. This suggests that although $Pnab$ and $A2_1am$ phases are close in energy, accessing the polar $A$2$_{1}am$ phase from the $Pnab$ ground state and realising antiferroelectric properties might require overcoming a significant energy barrier: this barrier might be due to not only "unwinding" the X$_{1}^{-}$ ($a^{0}a^{0}c^{-}$) tilts (analogous to the loss of the [001]$_{t}$ tilt in the low temperature -- intermediate temperature phase transition in Bi$_{2}$WO$_{6}$~\cite{Bi2WO6_Lightfoot}), but also removing the anti-polar M$_{5}^{+}$ displacements to give the possible intermediate $a^{-}a^{-}c^{0}$ phase of $Cmcm$ symmetry which is high in energy. By contrast, this hypothetical intermediate $Cmcm$ $a^{-}a^{-}c^{0}$ phase is relatively much lower in energy for both Sr$_{3}$Zr$_{2}$O$_{7}$ and Sr$_{3}$Sn$_{2}$O$_{7}$.~\cite{Sr3Zr2O7,Sr3Sn2O7}
It would be interesting to explore the effect of substitution of spherical cations such as La$^{3+}$ into the bismuth sites in the fluorite-like layers (whilst maintaining the Aurivillius structure) to see how the balance of energies of the $\Gamma_{5}^{-}$ or M$_{5}^{+}$ displacements and octahedral tilts can be tuned.

\section{Conclusions}

We have shown consistently from first-principles simulations and experimental work, that the ground state of the $n$=2 Aurivillius Bi$_2$W$_2$O$_9$ structure is an non-polar phase of $Pnab$ symmetry. This phase appears as a small distortion of the paraelectric $I4/mmm$ parent phase involving an in-plane antipolar displacement of W and Bi cations and in-plane and out-of-plane octahedral tilts $(a^-a^-c^-)$. Close in energy to this ground state, we identified a metastable polar phase of $A2_1am$ symmetry, involving an in-plane polar displacement of W and Bi cations and in-plane and out-of-plane octahedral tilts $(a^-a^-c^+)$. The energy proximity between these polar and antipolar phases, related by a first order phase transition,  makes  Bi$_2$W$_2$O$_9$ a potential antiferroelectric material. The electric field required to stabilize the $A2_1am$ phase against the $Pnab$ ground state was estimated theoretically to be 190 kV cm$^{-1}$. Experimentally, we confirm the non-ferroelectric character of Bi$_2$W$_2$O$_9$,  but applying fields up to $\pm$250 kV cm$^{-1}$ was not sufficient to reveal the antiferroelectric behavior. This should however be investigated by applying larger fields. The antiferroelectric character might also be favored and revealed by appropriate strain engineering of the energy landscape.


\section{Acknowledgements}

Computational resources are provided by the Consortium des Equipements de Calcul Intensif (CECI), funded by the F.R.S.-FNRS under the Grant No. 2.5020.11 and the Tier-1 supercomputer of the F\'ed\'eration Wallonie-Bruxelles funded by the Walloon Region under the Grant No 1117545.  H.D. and Ph.G. acknowledge support from Algerian-WBI bilateral cooperative project. We are grateful to the ISIS Neutron and Muon Source for provision of NPD time through the HRPDXpress scheme (DOI \textcolor{red}{10.5286/ISIS.E.RB1790213}) and to Dr A. Gibbs and Dr D. Fortes for their assistance. EB and EEM are grateful to the Royal Society (IES-R3-170112) and to the Leverhulme Trust (RPG-2017-362) for funding. WZ and PSH thank the Welch Foundation (grant E-1457) for support.

\bibliography{240419_bib}

\end{document}